\documentclass[runningheads]{llncs}

\usepackage{amsmath}

\usepackage[T1]{fontenc}
\usepackage{float}
\usepackage{amsfonts}
\usepackage{amssymb}
\usepackage{makecell}
\usepackage{graphicx}
\usepackage{bm}
\usepackage{algorithm}
\usepackage{algorithmic}

\begin{document}
\title{Attention-ResUNet and EfficientSASM-UNet: UNet based frameworks for Lung and Nodule segmentation}
\titlerunning{Attention-ResUNet and EfficientSASM-UNet} 

\author{
Muhammad Abdullah\inst{1} \and 
Furqan Shaukat\inst{1} 
}
\authorrunning{M. Abdullah et al.}
\institute{
Faculty of Electrical and Electronics Engineering, University of Engineering and Technology, Taxila, Pakistan \\
\email{furqan.shoukat@uettaxila.edu.pk}
}

\maketitle

\begin{abstract}
Lung cancer has been one of the major threats across the world with the highest mortalities. Computer-aided detection (CAD) can help in early detection and thus can help increase the survival rate. Accurate lung parenchyma segmentation (to include the juxta-pleural nodules) and lung nodule segmentation, the primary symptom of lung cancer, play a crucial role in the overall accuracy of the Lung CAD pipeline. Lung nodule segmentation is quite challenging because of the diverse nodule types and other inhibit structures present within the lung lobes. Traditional machine/deep learning methods suffer from generalization and robustness. Recent Vision Language Models/Foundation Models perform well on the anatomical level, but they suffer on fine-grained segmentation tasks, and their semi-automatic nature limits their effectiveness in real-time clinical scenarios. In this paper, we propose a novel, fully automatic method for accurate 3D segmentation of lung parenchyma and lung nodules. The proposed architecture is an attention-based, fully convolutional UNet with residual blocks at each encoder-decoder state. Dilated convolutions at each encoder-decoder stage allow the model to capture the larger context without increasing computational costs. For Lung Nodule Segmentation an Efficent-SASM(Self Attentive Similarity Module) based 3D-UNet is proposed which is optimized for easy deploymnet. The proposed methods have been evaluated extensively on one of the largest publicly available datasets, namely LUNA16, and is compared with recent notable work in the domain using standard performance metrics like Dice score, IOU, etc. It can be seen from the results that the proposed Lung Segmentation method achieves better performance than state-of-the-art methods, while comparable results are achieved by proposed Nodule Segmentation technique. The source code, datasets, and pre-processed data can be accessed using the link: https://github.com/EMeRALDsNRPU/Attention-Based-3D-ResUNet.

\keywords{Lung Nodule Segmentation  \and Lung Parenchyma Segmentation\and Computer Aided Detection \and Lung Cancer.}
\end{abstract}
\section{Introduction}
Lung cancer ranks among the most prevalent cancers globally, with around 2.2 million new cases recorded in 2020\cite{sung2021global}. The American Cancer Society estimated almost 234,580 new cases of lung cancer in the US for the year 2024 only. Approximately 340 people die because of lung cancer daily in the US, nearly 2.5 times the fatalities attributed to colorectal cancer (CRC), which ranks second in cancer mortality\cite{siegel2024cancer}. A survey by the European Society for Medical Oncology (ESMO)\cite{planchard2018metastatic}, suggests that the highest prevalence of lung cancer occurs in Central/Eastern European and Asian populations. The situation in developing countries (such as in Asia) is particularly dire\cite{moore2010cancer}. Research indicates that early detection of lung nodules can significantly improve the survival rate \cite{siegel2019cancer}. However, early detection of lung cancer is difficult primarily because of 1) the absence of early symptoms in the majority of patients. 2) a substantial volume of data available in the form of computed tomography (CT) scans, and 3) the interobserver variability in nodule detection. Computer-aided detection (CAD) can facilitate the early detection of lung nodules and decrease the workload of expert personnel. Lung nodules are typically characterized as the primary symptom of lung cancer, developing within and at the peripheries of the lungs. In radiology, CAD systems can help clinicians in medical image analysis\cite{shaukat2019computer} to detect and localize structures of interest in a semi-automatic or fully automated manner. The five-year survival rate for lung cancer patients is the lowest among all cancers, indicating a pressing need for the development of automated and reliable systems to enhance early detection, diagnosis, and treatment. 
Accurate segmentation of lung parenchyma \& lung nodules plays a crucial role in the overall accuracy of the lung CAD pipeline where accurate segmentation of lung lobes ensures the inclusion of juxta-pleural nodules. Lung Nodule segmentation has its own challenges because of the diverse nodule types and other inhibit structures present within the lung lobes. Extensive research has been conducted on the development of lung nodule segmentation systems utilizing traditional machine learning and deep learning methodologies\cite{zhang2023lung,poonkodi2023mscaunet,gang2025segmentation,lu2023weakly,bbosa2024mrunet,agnes2024wavelet,selvadass2024satunet,cai2024mdfn}. However, traditional machine learning or deep learning methods suffer from generalization and robustness. 

The recent emergence of massive visual language models (VLMs) and their ability to generalize to unseen tasks has prompted the healthcare community to adapt them for numerous medical applications. The Segment Anything Model (SAM) \cite{kirillov2023segment} has excelled in various segmentation tasks using standard real-world images. The domain gap between real-world and medical images, such as those in radiology, and its adaptability have been examined in several variants of SAM \cite{ma2024segment,qiu2023learnable,gong20233Dsam,shaharabany2023autosam,shaukat2024lung}. Nevertheless, these models have primarily been investigated at the anatomical level, while lesion-level detection remains a challenge because of the fine-grained segmentation nature. In addition, the semi-automatic nature of these VLMs limits their application in real-time clinical scenarios. 
To this end, we present a novel fully automatic method for accurate 3D segmentation of lung parenchyma and lung nodules.
\subsection{Contributions}
Our \textbf{contributions} in this work can be summarized as follows:
\begin{itemize}
\item Development of a fully automatic multi-scale attention network for automatic 3D segmentation of lung parenchyma and nodules.
\item Proposing a fully convolutional resUnet by replacing the max-pooling with strided convolutions at the encoder and trilinear interpolation by transposed convolutions at the decoder, which maximizes the number of learnable parameters.
\item Adapted dilated convolutions at each encoder-decoder stage which allows the model to capture the larger context without increasing any computational cost.
\item Increasing the depth of the residual block at each stage thus easing the process of learning the complex patterns and fine details in the dataset.
\item For lung nodule segmentation, an adaptive multi-head attention block is proposed, which automatically calculates the number of attention heads in each residual block.
\end{itemize}

\section{Methodology}
\subsection{Dataset curation and Preprocessing}

The LUNA dataset\cite{setio2017validation} is utilized for training and validation of the proposed model. LUNA is a subset of the LIDC-IDRI dataset\cite{armato2011lung} which contains 1018 CT (Computed Tomography) scans annotated by four expert radiologists in double-blinded fashion. In LUNA, the inconsistent cases have been removed from the original dataset resulting in a total of 888 scans. The nodule inclusion criteria of LUNA have been followed in subsequent nodules’ evaluations which gives a total of 1186 nodules. The corresponding lung masks are also given, which are used as labels for training the proposed model for lung segmentation. Nodule masks are taken from the original LIDC-IDRI dataset of respective cases, which are used as labels for training the proposed model for nodule segmentation.

Each scan of a 3D CT image and corresponding label is resized to 300×300 for lung segmentation and 64×64 for nodule segmentation. For lung lobe segmentation, after resizing, immediate 11 scans around each side of the median slice were taken, thus making a total of 23 slices (11+11+median slice). For lung nodule segmentation, 64 slices of nodule appearance were considered. This cropping was performed because most image information is contained around the median slice. In short, resizing and cropping help manage the variables in memory. 
\subsection{Network architecture for Lung segmentation}
\begin{figure}[H]
\includegraphics[width=\textwidth]{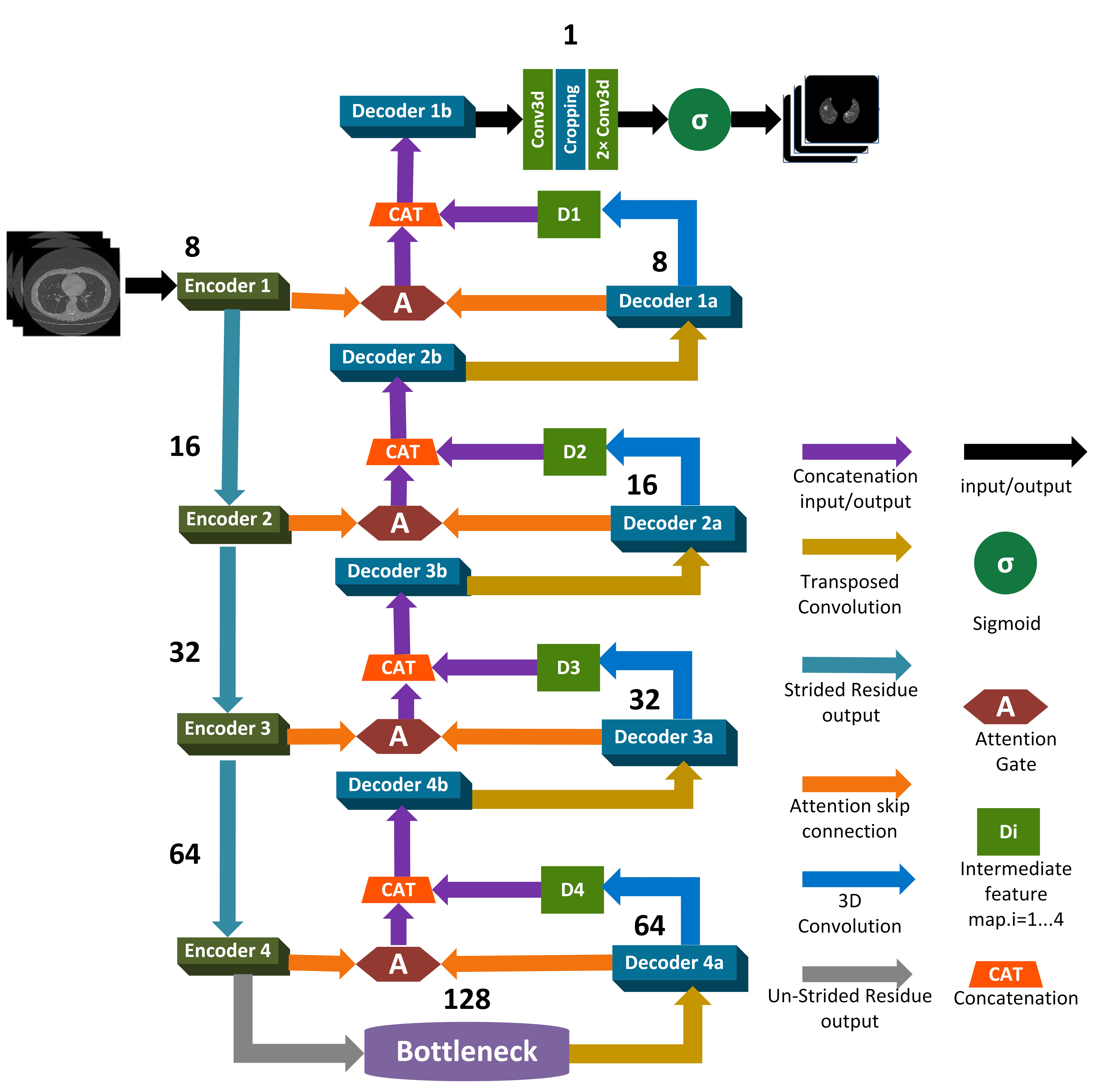}
\caption{Proposed Lung segmentation Network. } \label{fig1}
\end{figure}

We have taken UNet\cite{ronneberger2015u} as a backbone of our proposed model with 4 encoder and decoder stages as shown in Figure \ref{fig1}. A bottleneck is used after the last encoder stage and before the first decoder stage. Each encoder-decoder stage and bottleneck contains a Residual block, which is shown in Figure \ref{fig2}. Furthermore, Transposed convolution (stride=2) is utilized at each decoder stage for upsampling, which replaces the trilinear interpolation, thus increasing the number of learnable parameters.

Attention gates are utilized between the last encoder stage $(e4)$ and the first decoder stage$(d4)$. Similarly $(e3,d3)$, $(e2,d2)$, and $(e1,d1)$ pairs serve as inputs to the attention gates. The working of an attention gate is shown in Figure \ref{fig3}. The result of the attention gate at each stage is concatenated with the decoder output after passing the output from a 3D convolution. The Concatenation operation is defined as:
\begin{figure}[H]
\includegraphics[width=\textwidth]{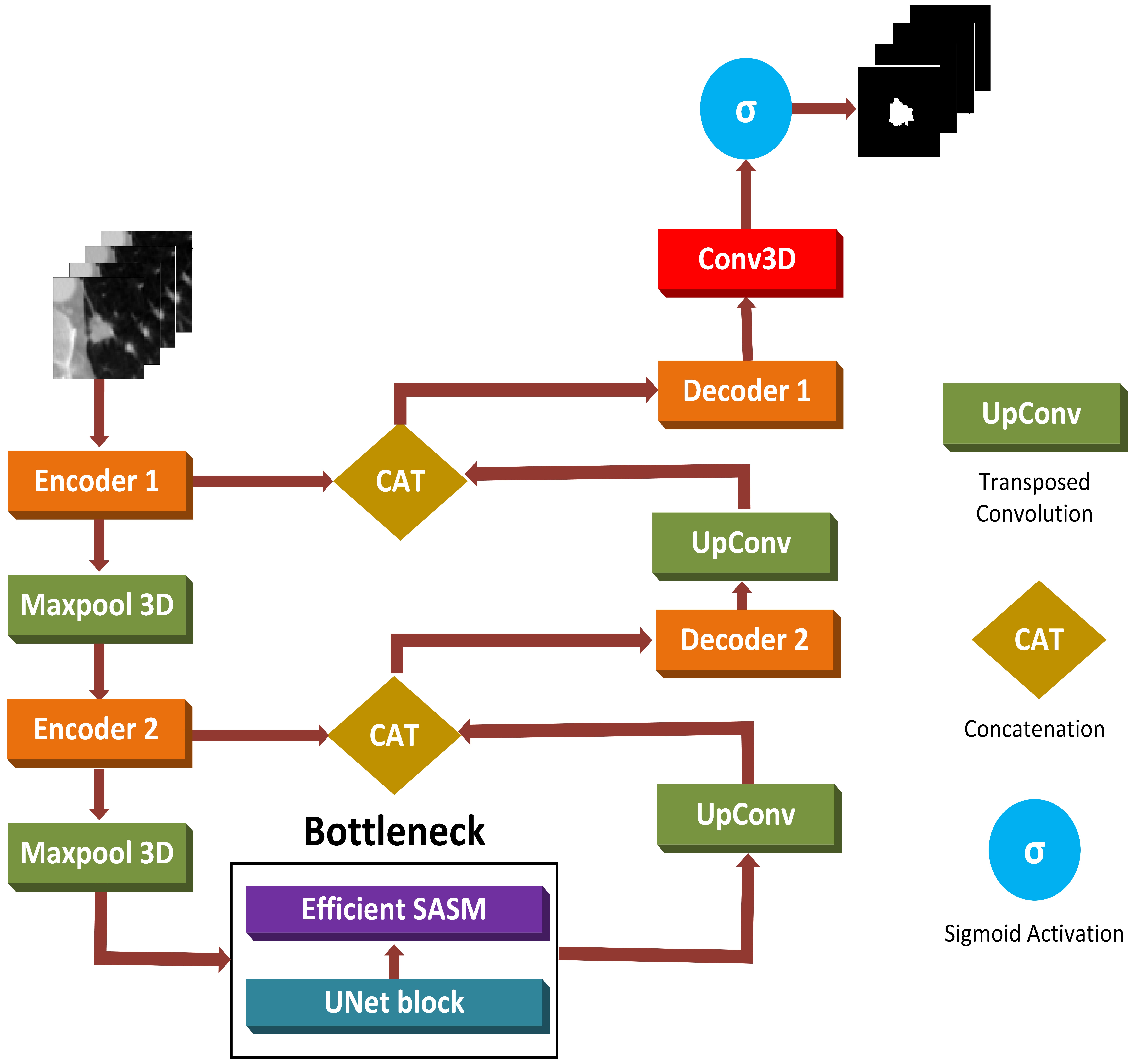}
\caption{Proposed Nodule segmentation Network.} \label{fig2}
\end{figure}
\begin{equation}
\mathrm{CAT} = \operatorname{cat}(A, D_i), \quad \text{where } i = 4, \dots, 1
\end{equation}

where $A$ represents the attention gate. $D_i$ is the output of the decoder after passing through 3D convolution.

The result of concatenation passes to the decoder b for further processing of concatenated features. The output features of decoder b pass to next decoder as input after upsampling by transposed convolution. The final output after the last concatenation operation passes through a 3D convolution and a cropping operation. Similarly, two additional convolutional operations are applied to mitigate the effect of cropping. Finally, the sigmoid function is applied to predict the binary output.

\subsubsection{Residual block}
Our proposed residual block is inspired by \cite{he2016deep} with certain modifications. The modified 3D Residual block contains 2 layers of operations, with each layer containing 3×3, 3D convolution (with stride 2 at the encoder and stride 1 at the decoder), and 3D Batch normalization. Sride 2 convolution replaces the maxpooling operation, thus increasing the number of learnable parameters. Similarly, dilation of 2 is utilized at each 3D convolution. 
 \begin{figure}[H]
\includegraphics[width=\textwidth]{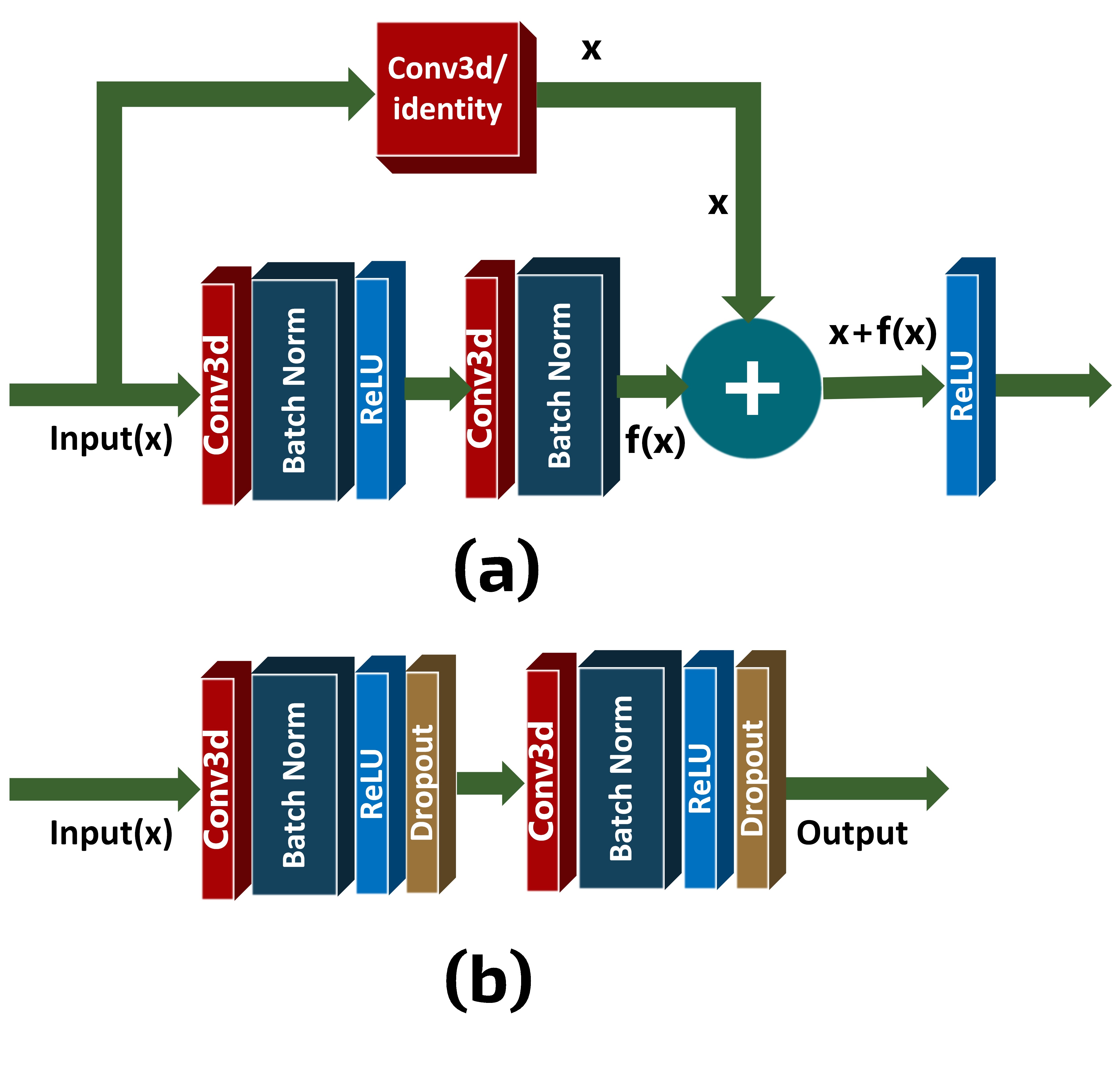}
\caption{Proposed 3D residual block with dilated convolutions.} \label{fig3}
\end{figure}
The dilated convolutions proposed in \cite{yu2015multi}, which are applied at each stage, expand the receptive field of the deep network without increasing the computational cost. Batch normalization helps in the fast convergence of the model while training. ReLU activation is applied after the first layer and the skip connection, which brings non-linearity to the model. Increasing the depth of the residual block helps in learning the fine details and complex patterns in the dataset. The output of the residual block is shown as:
\begin{equation}
\mathrm{Out} = \sigma\bigl(x + f(x)\bigr)
\end{equation}

where $x$ is input. $f(x)$ is output after two layers of residual block. $\sigma$ is ReLU activation.

\subsubsection{Attention Gates}
The concept of an attention gate proposed by \cite{oktay2018attention} is extended to 3D in our proposed scheme. Three-dimensional (3D) attention gates with dilated convolutions (i.e., dilation = 2) used in our proposed scheme are shown in Figure 3. Two additional 3D transposed convolutions (stride =2) and 3D convolutions were added before the addition and multiplication of feature maps to bring depth in attention gates and hence more robust features are learnt. The output $z_0$ is a 1-channel feature map, which is multiplied with the transformed input $f_t(x)$ to get the attention output $z$. The input signal is the encoder output, and the gating signal is the decoder output of each stage.
 \begin{figure}[H]
\includegraphics[width=\textwidth]{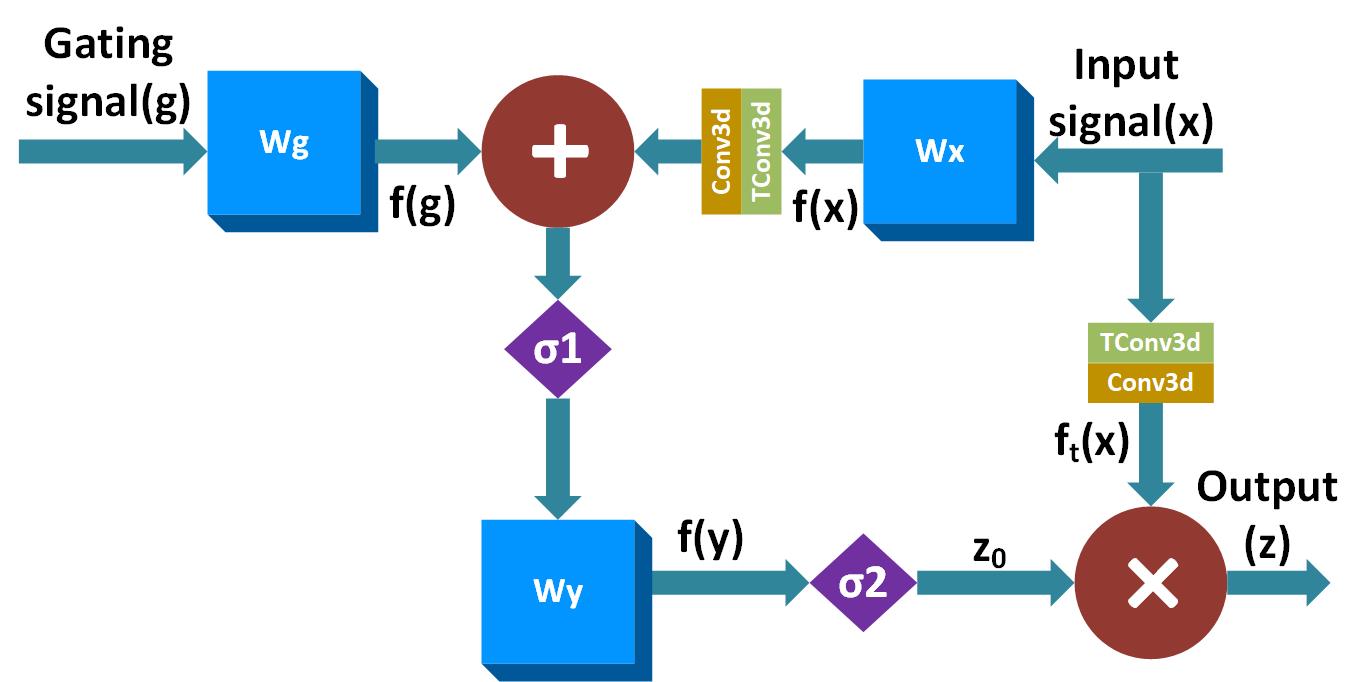}
\caption{Proposed 3D Attention Gate with dilated convolutions. Dilated convolution operations are shown as $W_x$, $W_g$ and $W_y$. ${\sigma}_1$ and ${\sigma}_2 $ show the ReLU and Sigmoid activations, respectively. Addition and multiplication of feature maps are shown by  $+$ and $\times$ respectively. TConv3D and Conv3D are transposed 3D strided convolutions and 3D  convolutions, respectively.} \label{fig4}
\end{figure}

\subsection{Network architecture for Nodule segmentation}
For nodule segmentation extra cropping layers were removed as shown in \ref{fig2} which can cause the model to memorize the patterns thus decreasing the efficiency and generalization of the proposed architecture. At each encoder stage maxpooling was utilized which decreases the number of learnable parameters to make the model simple for deployment. Similarly, decoder A and B stages were were replaced with only one decoder at each stage to prevent the overfitting which could arise due to complex nature of the network. Attention Gates were also removed to make the model as simple as possible. Similary, Efficient Self Attentive Similarity Attention Module(SASM) was proposed and utilized in the bottleneck.
 \begin{figure}[H]
\includegraphics[width=\textwidth]{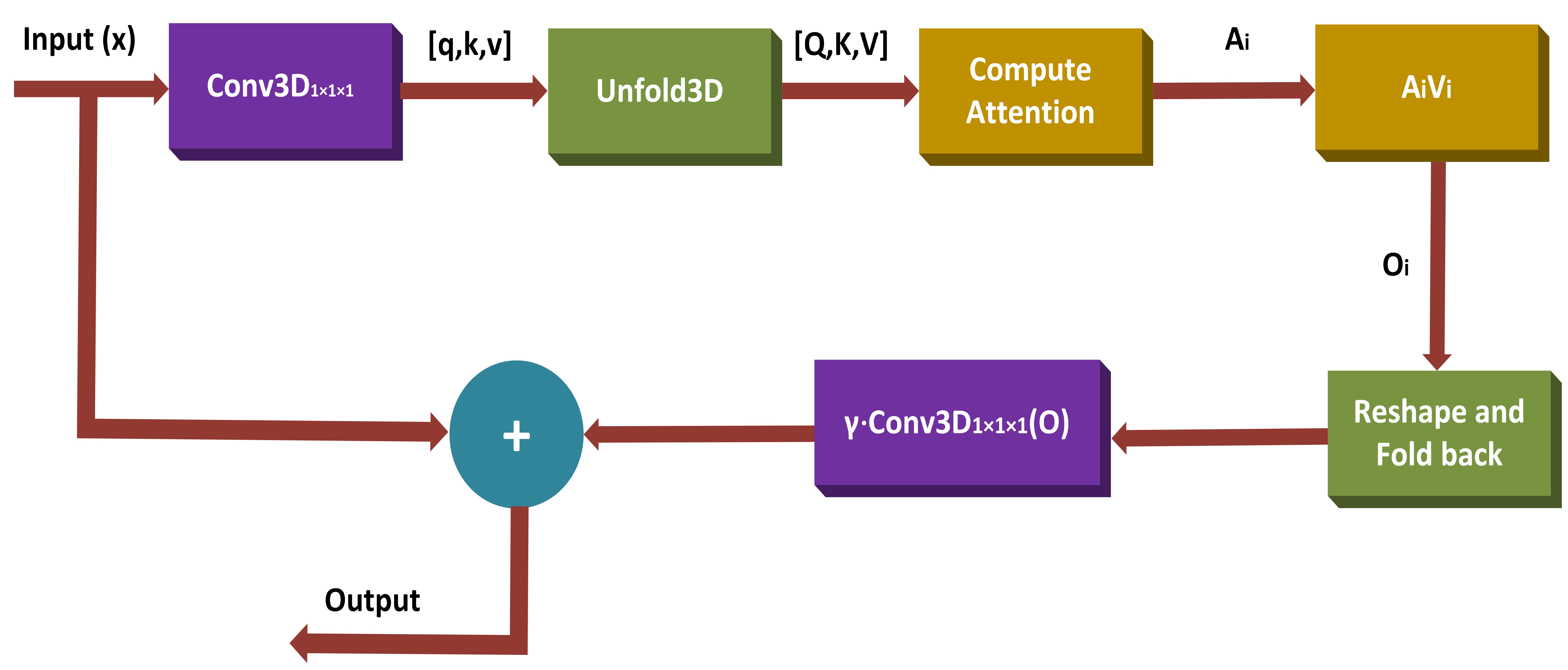}
\caption{Proposed Efficient SASM module} \label{fig5}
\end{figure}

\subsubsection{Efficient SASM}

The proposed Efficient SASM module is shown in \ref{fig5}
\begin{algorithm}[H]
\caption{EfficientSASM: Efficient Self-Attention Similarity module}
\begin{algorithmic}[1]
\REQUIRE Input tensor $x \in \mathbb{R}^{B \times C \times D \times H \times W}$, window size $W_s = (w_d, w_h, w_w)$
\STATE Compute QKV tensor using $1 \times 1 \times 1$ 3D convolution: $\text{QKV} \leftarrow \text{Conv3D}{1 \times 1 \times 1}(x)$
\STATE Split QKV into query, key, value: $q, k, v \leftarrow \text{Split}(\text{QKV})$
\STATE Unfold $q, k, v$ into non-overlapping 3D windows of size $W_s$
\STATE Compute tokens per window: $N_w \leftarrow w_d \cdot w_h \cdot w_w$
\STATE Compute total number of windows: $N \leftarrow \frac{D}{w_d} \cdot \frac{H}{w_h} \cdot \frac{W}{w_w}$
\STATE Reshape $q, k, v \in \mathbb{R}^{B \times N \times N_w \times C}$
\STATE Compute attention: $A \leftarrow \text{Softmax}\left(\frac{q \cdot k^\top}{\sqrt{C}}\right)$
\STATE Apply attention to values: $o \leftarrow A \cdot v$
\STATE Fold back into 3D volume: $o' \leftarrow \text{Fold}(o)$
\STATE Final projection: $\hat{o} \leftarrow \text{Conv3D}{1 \times 1 \times 1}(o')$
\STATE Output with residual: $\text{Output} \leftarrow x + \gamma \cdot \hat{o}$, where $\gamma \in \mathbb{R}$ is learnable
\RETURN $\text{Output} \in \mathbb{R}^{B \times C \times D \times H \times W}$
\end{algorithmic}
\end{algorithm}

Efficient SASM module is similar to transformers with a number of key differences. It unfolds local 3D windows and applies attention to local cubes instead of flattening the non-overlapping pateches and using patch embedding and positional encodings. Instead of global self attention it uses windowed self attention and utilizes 1$\times$1$\times$1 convolution to compute Q,K and V instead of using the dense layer. More efficient than transformers for 3D volumatric images where global attention is not feasible. Instead of linear MLP layers it uses 1$\times$1$\times$1 convolutional layer for QKV projection. Unfolding refers to the process of dividing larger 3D windows into smaller 3D windows i.e 4$\times$4$\times$4 window is divided into 8, 2$\times$2$\times$2 windows and Folding does the reverse process. The smaller unfolded windows require less computation and is therefore more efficient for 3D segmentation.
\subsubsection{Residual blocks replaced by simple Convolutional Blocks}
Input is fed to 3D convolution blocks followed by Batch normalization and ReLU activation. Skip connections were removed to optimize the model for better performance at runtime. Two dropout layers are utilized after each of two RelU activation as shown in \ref{fig3}, to prevent overfitting while training the model.
\subsection{Loss function}
The loss function utilized for lung and nodule segmentation is a combination of Binary cross-entropy and Dice loss. Binary cross-entropy and Dice loss are robust in nature and are widely adapted in medical image segmentation tasks.
Binary cross entropy is defined as:
\begin{equation}
L_{BCE} = - \frac{1}{N} \sum_{i=1}^{N} [m_i logp_i + (1-m_i)log(1-p_i)],
\end{equation}

Similarly, Dice loss is defined as:
\begin{equation}
L_{Dice} = 1 - \frac{2\sum_{i=1}^{N}m_ip_i}{\sum_{i=1}^{N}(m_i)^2+\sum_{i=1}^{N}(p_i)^2} ,   
\end{equation}

The final loss L is defined by 
\begin{equation}
L = L_{BCE} + L_{Dice},
\end{equation}
\section{Experimental Results and Discussion.}
The LUNA 16\cite{setio2017validation} dataset was used for training and validation purposes.The proposed networks was trained for lung parenchyma segmentation and nodule segmentation. A total of 100 epochs were used in both cases. The proposed models was evaluated for their respective tasks, i.e., lung parenchyma segmentation , using standard performance metrics like Dice Score, IOU, Precision, and Recall, and was compared with other notable work in the domain.
\subsection{Lung segmentation results}
 The dataset was split into the train, validation, and test sets with a ratio of 60-20-20, respectively for lung segmentation.  A batch size of 01 was selected due to the limited memory used to handle large arrays. The loss function was minimized using the Adam optimizer with a learning rate of 1e-4. A small learning rate improves the generalizability of the model and helps in learning complex patterns in the dataset. The trained model was tested on the 20\%Table  test dataset, which was never seen by the model  \ref{tab1} which represents lung parenchyma segmentation results. We have tried to include the state of the art and recent studies in our comparison. For lung parenchyma segmentation, the test results of the recent studies \cite{zhang2023lung,poonkodi2023mscaunet,gang2025segmentation} and results of SOTA like UNet, AttentionUNet, and DeepLabV1 reported by \cite{gang2025segmentation}, along with ours, are shown in Table 1 for comparison.
 \begin{table}[t]
\centering
\caption{Comparison of Lung Parenchyma Segmentation Results with Existing Methods.}\label{tab1}
\begin{tabular}{l| c| c |c |c |c}
\hline
\textbf{Author}	
&\textbf{Year} 
&\makecell{\textbf{Method}}
&\makecell{\textbf{No. of} \\\textbf{samples}}	
&\makecell{\textbf{Dice Score}\\ \textbf{(\%)}}	
&\makecell{\textbf{IoU(\%)}}\\
\hline

Ronneberger et al. \cite{ronneberger2015u} &2015	&\makecell{UNet}	&\makecell{163 CT Scans}	&91.52 	&84.73 	 \\

Chen et al. \cite{chen2017deeplab} &2017	&\makecell{DeepLabV1}	&\makecell{163 CT Scans}	&88.67 	&80.77  \\

Oktay et al. \cite{oktay2018attention} &2018	&\makecell{Attention-UNet}	&\makecell{163 CT Scans}	&91.96 	&86.48  \\

Zhang et al.  \cite{zhang2023lung} &2023	&\makecell{NASNet-large \\ Net}	&\makecell{1000 X-Ray \\ Images}	&92.00	&87.00	\\

Poonkodi et al. \cite{poonkodi2023mscaunet} &2023	&\makecell{MSCAUNet-3D}	&\makecell{335 CT Scans}	&91.40  	&90.40	\\

Gang et al. \cite{gang2025segmentation} &2025	&\makecell{Pyramidal \\Pooling Channel \\Transformer\\ (PPCT)}	&\makecell{163 CT Scans}	&88.99 	&80.40 	 \\
Li et al. \cite{li2025ae} &2025	&\makecell{AE-UNet}	&\makecell{2729 CT images}	&72.22 	& 57.15 	 \\
 
\textbf{Ours} &2025	&\makecell{3D-Attention \\ResUNet}	&\makecell{888 CT Scans}	&\textbf{93.61}	&\textbf{91.03}	 \\
\hline
\end{tabular}
\end{table}

 It can be seen that our proposed method outperforms the other studies in lung segmentation tasks, even though our method is evaluated on a considerably large and well-established dataset as compared to other techniques. It can be seen that the model in \ref{fig1} performs quite well on Lung Segmentation. Some other comparable works with respect to architecture for this task are \cite{maji2022attention,agarwala2022unet,alshomrani2023saa,bbosa2024mrunet,yu2022ags}. However, we have added the following incremental novelties to our proposed architecture to improve its performance and generalization ability. 3D attention gates and residual blocks have been used compared to 2D used in \cite{maji2022attention}. Similarly, the proposed architecture uses 3D fully convolutional layers instead of max pooling and 3×3 transposed convolutions at the decoder stage. Agarwala et al.\cite{agarwala2022unet} uses 3D UNet,  but single encoded-decoder blocks utilize simple 3D convolutional blocks without Residual connections. Our proposed Residual block not only utilizes skip connections but has 2× increased depth. SAA-UNet \cite{alshomrani2023saa} proposed by Alshomrani et al. has a similar 2D architecture, but residual connections were not utilized at single encoder-decoder stages. MRUNet-3D\cite{bbosa2024mrunet} is also a comparable architecture but lacks the attention gates. Our model also replaces the trilinear interpolation with transposed convolutions, which increases the number of learnable parameters. AGs-UNet, proposed by Yu et al.  \cite{yu2022ags}, lacks batch normalization at each stage, which is essential for fast convergence. Furthermore, our proposed model utilizes the dilated convolution at each stage, which expands the receptive field of the deep network without increasing any computational cost. An expanded receptive field allows the model to capture a broader context in an image.
 \subsection{Nodule segmentation results}
 Similarly, for lung nodule segmentation, test results of the recent studies\cite{zhang2023lung,poonkodi2023mscaunet,gang2025segmentation} along with ours are shown in \ref{tab2} for comparison. Would like to note that a 3D crop of the size $64\times64\times64$ of a CT scan was used as input to the nodule segmentation model. As this cropping decreases the context for the 3D nodule, a fall in dice score and IoU is observed. These results can be further improved by considering the larger cropped image size, which will provide more information around the lung nodule and help the model better understand the CT image in a broader context. Data augmentation techniques can also be utilized to improve the performance of the model. Similarly, the Efficient-SASM module can be inserted in encoder and decoder as well instead of only in the bottleneck which will ensure the better selection of encoder-decoder features.

 \subsection{Qualitative Analysis}
\begin{figure}
\includegraphics[width=\textwidth]{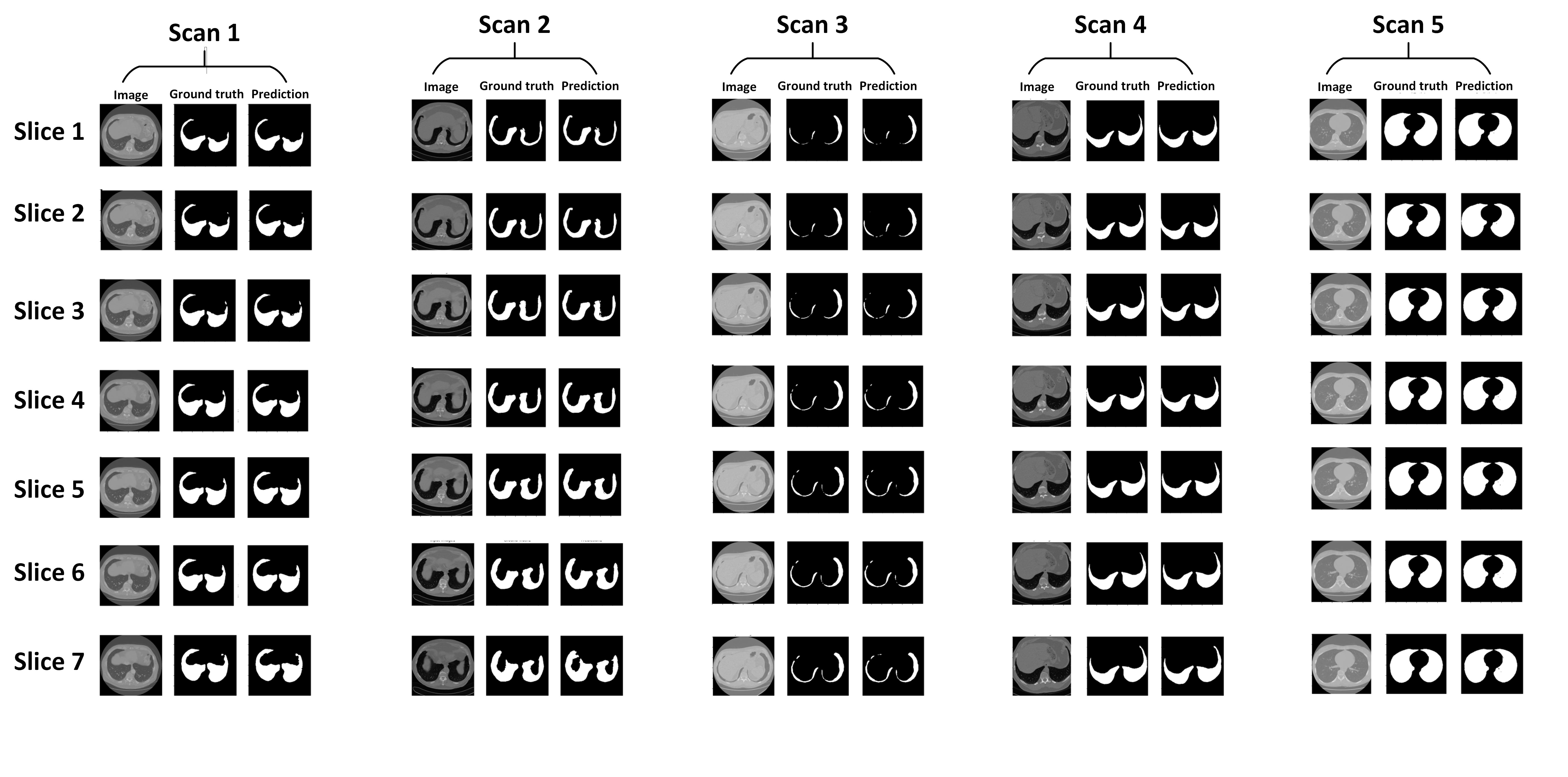}
\caption{Sample results of lung parenchyma segmentation. The first column of each scan represents different slices, the second column represents the ground truth and third column represents the segmentation results} \label{fig6}
\end{figure}

\begin{figure}
\includegraphics[width=\textwidth]{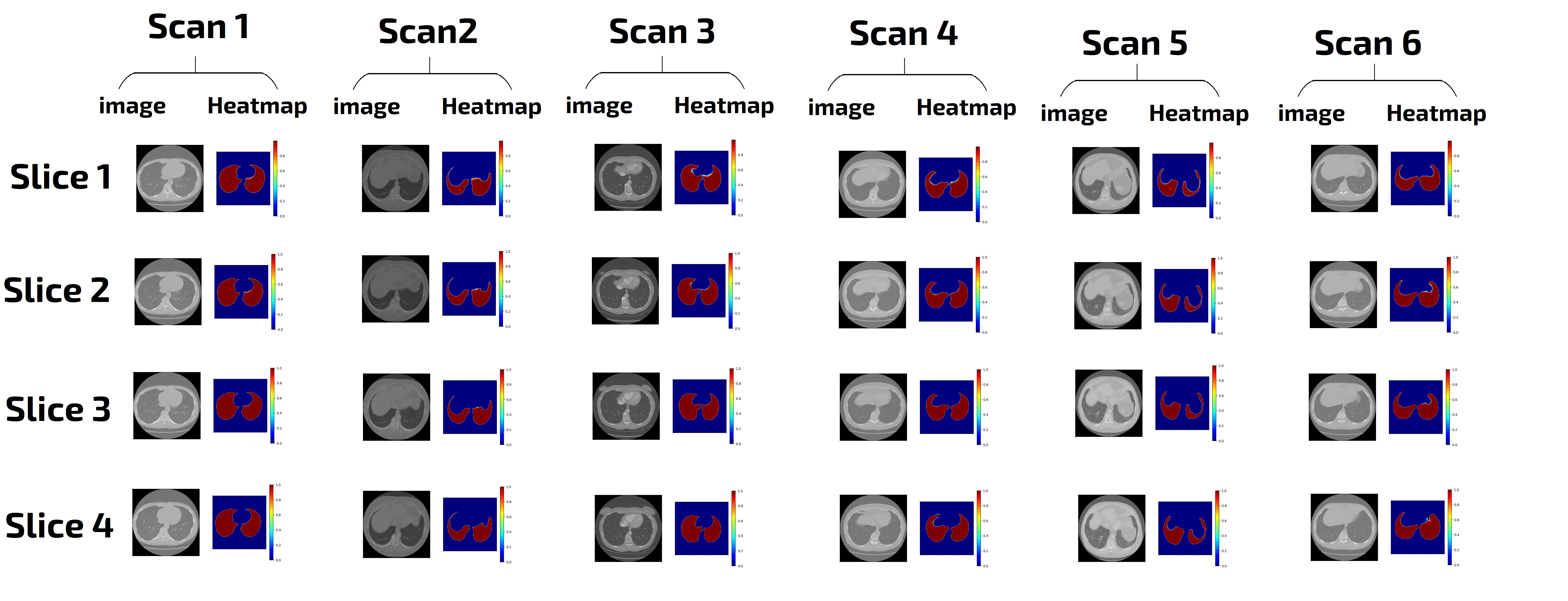}
\caption{Probability Heat maps of Attention ResUNet showing pixel probability of predicted logits} \label{fig7}
\end{figure}

\begin{figure}[H]
\includegraphics[width=\textwidth]{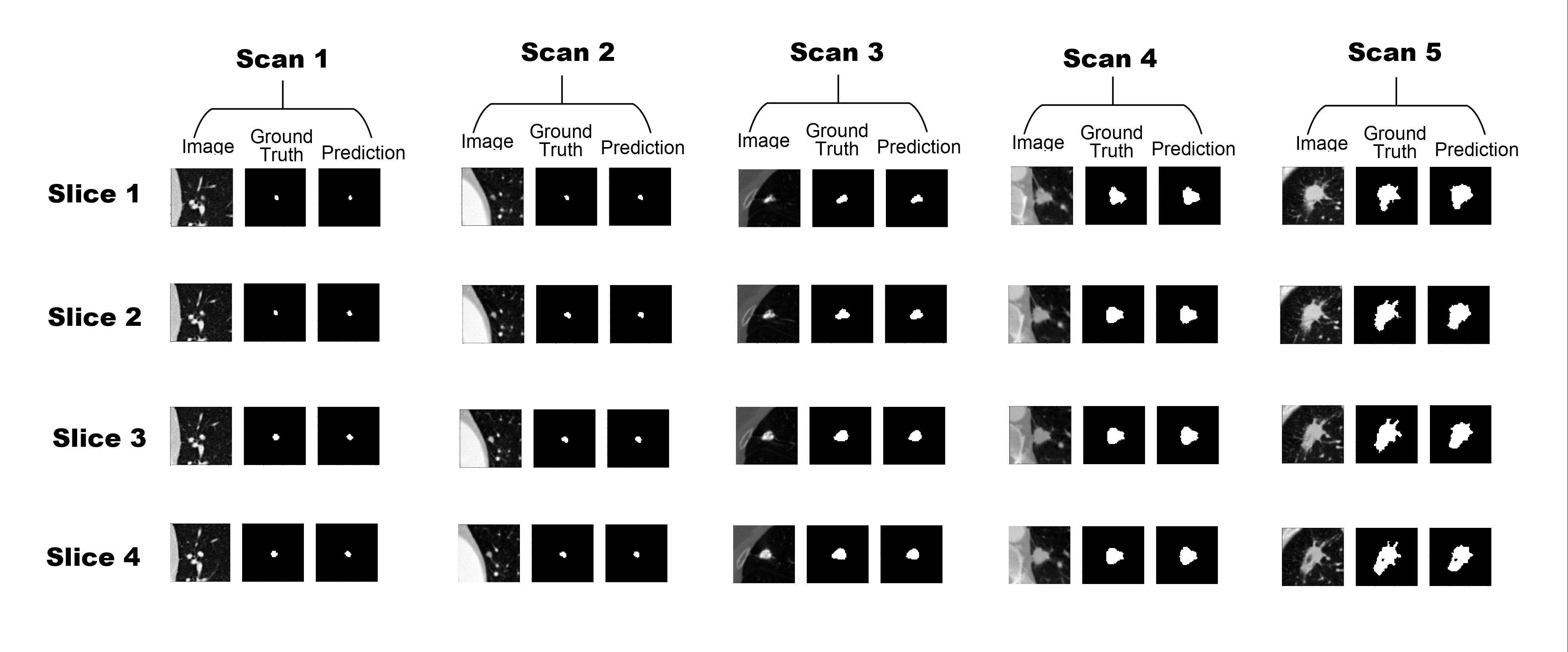}
\caption{Sample results of lung nodule segmentation. The first column of each scan represents different slices, the second column represents the ground truth and third column represents the segmentation results} \label{fig8}
\end{figure}

\begin{figure}[H]
\includegraphics[width=\textwidth]{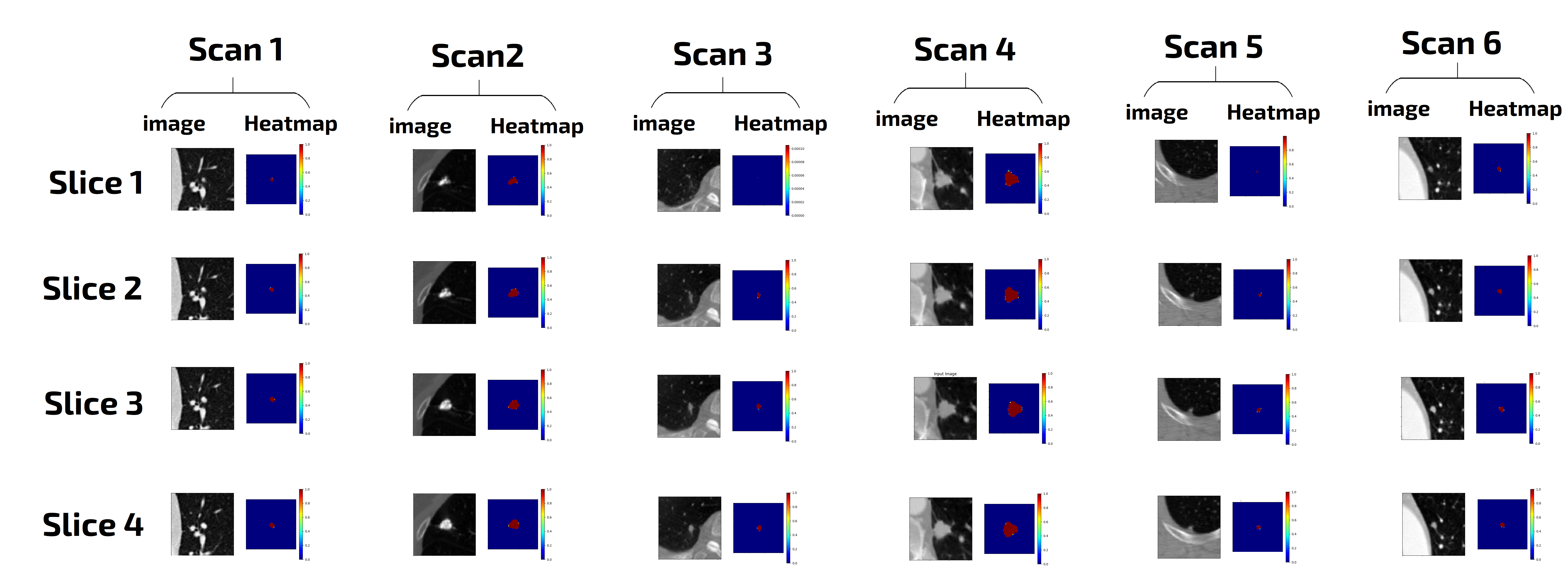}
\caption{Probability Heat maps of Efficient SASM UNet showing pixel probability of predicted logits} \label{fig9}
\end{figure}

 For qualitative comparison, we have included the sample results of both tasks as shown in \ref{fig6} and \ref{fig8}, respectively. \ref{fig6} shows the inference results of the proposed model for lung parenchyma segmentation, and \ref{fig8} shows the inference results for lung nodule segmentation. Lung segmentation captures the fine details and ensures precise segmentation of boundaries and edges. Nodule segmentation detects small, medium and large nodules accurately whereas it also ensures the correct construction of the shape of the nodule mask. \ref{fig7} and \ref{fig9} show the probability heat maps of lung parenchyma and lung nodule segmentation respectively. The dominating maroon/red color in foreground shows that most of the pixels are segmented correctly.The lung portion is clearly highlighted in \ref{fig7} and nodules are also identified in red color in \ref{fig9}. In other words maroon/red pixels show true positives and blue pixels show true negatives or background which in case of lung segmentation represents air or fluid around the lung and in case of nodule segmentation represents all other areas of lung around the nodule. The accurate separation of these pixels shows improved performance.  Other color pixels show some false positives or false negatives which are very small in number.


\begin{table}[t]
\centering
\caption{Comparison of Lung Nodule Segmentation Results with Existing Methods.}\label{tab2}
\begin{tabular}{l| c |c| c| c| c}
\hline
\textbf{Author}	&\textbf{Year} 
&\makecell{\textbf{Method}}	
&\makecell{\textbf{No. of} \\\textbf{samples}}	
&\makecell{\textbf{Dice Score}\\\textbf{(\%)}}	
&\makecell{\textbf{IoU (\%)}} \\
\hline
Bbosa et al.\cite{bbosa2024mrunet} &2024	&\makecell{MRUNet-3D} &\makecell{888 CT Scans}	&83.47 	&86.27 \\
Agnes et al.\cite{agnes2024wavelet} &2024	&\makecell{Wavelet U-Net++}	&\makecell{1018 CT Scans}	&93.70	&87.80 \\
Selvadass et al.\cite{selvadass2024satunet} &2024	&\makecell{SAtUNet} &\makecell{1018 CT Scans}	&81.10	&72.24 \\
Cai et al.\cite{cai2024mdfn} &2024	&\makecell{MDFN: A \\Multi-level Dynamic \\Fusion Network}	&\makecell{888 CT Scans}	&89.19	&80.78 \\
\textbf{Ours}	&2025 &\makecell{3D-EfficientSASM\\ -UNet}	&\makecell{888 CT Scans} 	&83.45	&72.12  \\
\hline
\end{tabular}
\end{table}

\section{Conclusions}

In this paper, we present a fully automatic 3D AttentionResUNet for lung parenchyma and EfficeintSASM-UNet for lung nodule segmentation. The proposed lung segmentation architecture is attention-based fully convolutional UNet with residual blocks at each encoder-decoder state. Maxpooling is replaced by strided convolutions at the encoder and trilinear interpolation is replaced by transposed convolutions at the decoder to maximize the number of learnable parameters. Dilated convolutions at each encoder-decoder stage allow the model to capture the larger context without increasing computational costs. Increasing the depth of residual block at each stage helps in learning the complex patterns and fine details in the dataset. One of the largest publicly available datasets, LUNA16, is used for model training and validation. The proposed method is compared with state-of-the-art methods using the standard performance metrics, exhibiting superior performance as compared to its counterparts. Efficent-SASM-UNet contains Efficient Self Attentive Similarity Module. The proposed method can be integrated into an end-to-end pipeline for lung nodule detection and its subsequent characterization. Both of the proposed methods can be used for early lung cancer screening and assist the medical community in the early diagnosis of lung cancer.

\begin{credits}
\subsubsection{\ackname} This work is part of the NRPU project $\#$ 17019 entitled “EMeRALDS: 
Electronic Medical Records driven Automated Lung nodule Detection and cancer risk 
Stratification” funded by Higher Education Commission of Pakistan.

\subsubsection{\discintname}
 The authors have no competing interests to declare that are
relevant to the content of this article. 
\end{credits}
%
%
%
\bibliographystyle{splncs04}
\bibliography{manuscript}

\end{document}